\documentclass[aps,pra,twocolumn,groupedaddress,showpacs,english]{revtex4}
\usepackage[T1]{fontenc}
\usepackage[latin9]{inputenc}
\usepackage{amsmath}
\usepackage{amssymb}
\usepackage{graphicx}

\usepackage{subcaption}
\usepackage{wasysym}
\usepackage[version=4]{mhchem}
\usepackage{collcell}


\usepackage{babel}

\begin{document}
\title{Three-dimensional-subwavelength field localization, time reversal of sources, and infinite-asymptotic degeneracy in spherical structures}

\author{Asaf Farhi}
\date{\today}
\begin{abstract}
High-resolution field localization in three dimensions is one of the main challenges in optics and has immense importance in fields such as chemistry, biology, and medicine. 
Time-reversal symmetry of waves has been a fertile ground for applications such as generating a subwavelength focal spot and coherent-perfect absorption. However, in order to generate the time reversed signal of a monochromatic source \emph{discrete} sources that are modulated according to the wave amplitude on a spherical envelope are required, rendering it applicable only in acoustics.
Here we approach these challenges by introducing a spherical layer with a resonant permittivity, which naturally generates the spatially \emph{continuous} time-reversed signal of an atomic and molecular multipole transition at the origin.
We start by utilizing a spherical layer with a resonant TM $l=1$ permittivity situated in a uniform medium to generate a free-space-subwavelength focal spot at the origin. We remove the degeneracy of the eigenfunctions of the composite medium by situating a point current source (or polarization) directed parallel to the spherical layer, which generates a focal spot at the origin \emph{independently} of its location. The free-space focal spot has a full width at half maximum of $0.4\lambda$ in the lateral axes and $0.58\lambda$ in the axial axis, which is tighter by a factor of $\sqrt{2}$ in each dimension in excitation-collection mode, overcoming the $\lambda/2$ far-field resolution limit in three dimensions.
We then explore two directions to localize electric field with deep-subwavelength resolution in three dimensions using this setup. 
Since the imaginary part of the eigenvalue is also realized in the physical parameter and the setup can be in an exact resonance, it can also open avenues in fields such as cavity QED, entanglement, and quantum information. In addition, we show that spherical structures exhibit a new type of degeneracy in which an infinite number of eigenvalues asymptotically coalesce. This high degeneracy results in a variety of optical phenomena such as strong scattering and enhancement of absorption and emission from an atom or molecule by orders of magnitude compared with a standard resonance.

\end{abstract}

\maketitle
The focal spot size that can be achieved by uniformly illuminating a circular aperture in the scalar approximation is given by an Airy disc, which is the Fourier Transform of a circular window \cite{jackson2012classical}. The full width at half maximum (FWHM) of this function is $1.02\lambda/\mathrm{NA},$ where
$\lambda$ is the wavelength and $\mathrm{NA}$ is the numerical aperture
$\left(\mathrm{NA}\lesssim1\right).$ This size is associated
with the lateral axes and in the axial axis the FWHM is $~2.5-3$ times larger due to the fact that a smaller range of $k_z\mathrm{s}$ is involved. For gaussian
beams, however, the focal spot is larger and depends on the width
of the beam. The optimal lens resolution enables to image most biological
cells but not viruses, proteins, and smaller molecules. Techniques such as confocal microscopy, structured illumination, beam shaping, and hyperlens imaging have been used to increase the lateral resolution \cite{minsky1988memoir,muller2010image,rho2010spherical}. In a $4\pi$
microscope the sample is illuminated from both sides and better resolution in the axial axis can be achieved \cite{cremer1974considerations}. However, in this setup side lobes are generated and the optical system needs to be realigned before every measurement in order for the focal spots to merge. Techniques based
on fluorescence such as STED \cite{hell1994breaking},
and PALM and STORM \cite{betzig2006imaging,moerner1989optical} enable
subwavelength resolution by stimulating emission at another frequency
using an additional torus-like illumination and by activating subsets of fluorescent molecules, which enables to accurately calculate the molecule locations, respectively. 
Maxwell fisheye is a spherical lens with a radius-dependent refraction index in which all light rays emitted
from a point meet at the antipodal point. The possibility of obtaining subwavelength resolution inside this setup has been the subject of recent works \cite {bitton2018two,alonso2015maxwell}. Time reversal of waves has also been applied for generating a subwavelength focal spot \cite{ScienceFink,ma2018towards}.
Finally, methods based on evanescent waves to enhance resolution such as near-field imaging and negative-refractive index lens enable subwavelength focusing usually for two-dimensional imaging \cite{pendry2000negative}. 
Here, we utilize a resonant spherical layer to localize far-field light in several settings. We first situate the spherical layer in a uniform medium and excite it with a point current. This setup generates a three-dimensional free-space subwavelength focal that has very minor side lobes.  Since it is composed of one "lens" it may not need to be aligned. In excitation-collection mode the effective focal spot is further minimized and there are almost no side lobes. We then explore two directions to localize far-field with deep-subwavelength resolution using this setup.

Time reversal of waves has been utilized for various interesting applications such as wave localization \cite{fink1992time,ScienceFink,lerosey2004time,ma2018towards} and coherent-perfect absorption \cite{CPA2012}. Recently, it was shown that the time reversal of a source, in the presence of a near-perfect absorber, results in a subwavelength focal spot \cite{ma2018towards}. In order to generate the time reversal of a wave generated by a source, one would have to let the wave propagate from the source, "freeze" time, and generate discrete sources on a spherical envelope modulated according to the wave amplitude. Here, we will utilize the resonant-spherical layer setup to generate the spatially-continuous time reversal wave of sources, enabling its use in electrodynamics.

Degeneracies of eigenvalues can arise from a symmetry of the system or from a special feature of the system. While the first type of degeneracy is widely known (e.g., $m$ degeneracy in spherical multipoles), the second, called accidental degeneracy, is more exotic and includes phenomena such as Landau levels \cite{landau1977quantum}, exceptional points \cite{bender1998real, el2007theory,ruter2010observation, lin2016enhanced,sweeney2019perfectly} and the accumulation point of the eigenpermittivities of evanescent modes \cite{bergman1979}.  
Degeneracies are associated with a strong response of the system as several modes are excited. In exceptional points for example, the degeneracy is usually second or third order and can lead to enhancement of emission from a molecule by two orders of magnitude due to enhancement of the density of states \cite{pick2017general}. In addition, the accumulation point of the eigenpermittivities of the evanescent modes can enhance the field (and emission) for a source that is very close to a metal-dielectric interface. Here we will show analytically and numerically that a spherical structure with a radius larger than $20\lambda$ exhibits infinite asymptotic all-even and all-odd TE/TM degeneracies of the second type. These degeneracies are associated with \emph{far field} and \emph{dielectric} spherical structures, in some cases with gain.

In a homogeneous medium the continuous-wave source-free solutions of Maxwell's equation are plane waves, vector spherical harmonics, and vector cylindrical harmonics.
It was recently shown that similarly to the situation in phased arrays
in which plane currents proportional to a homogeneous medium source-free solution
with a planar geometry generate the same function, currents proportional
to a vector spherical harmonic (VSH) on a spherical surface generate
the same VSH. Interestingly, a TM $l=1$ VSH near the origin has a subwavelength
far-field focal spot \cite{farhi2017generating}, which is smaller in volume by a factor of $\sim27$
compared with the focal spot that can be achieved by uniformly illuminating a lens. For a medium with a refractive index larger than 1, the
TM $l=1$ field will have even a smaller focal spot. Generating this mode by oscillating currents can be thought of as a continuous-wave time reversal of the field of an oscillating dipole at the origin. Importantly, generating these VSH propagating towards the origin are the time reversal of the atomic and molecular multipole transitions. It is thus of interest to generate these modes near the origin.  However, the
spatial distributions of these VSH are complex and a setup of currents
modulated accordingly is infeasible. 


Eigenfunctions of Maxwell's equations are fields, which exist without a source for certain physical parameters that correspond to resonances of the system \cite{bergman1980theory,ge2010steady}. Here, 
we utilize resonances in a setup of a spherical layer in a host medium to naturally generate a the
VSHs. This setup requires only a point source or emission in order to generate these field patterns. The permittivity value of the spherical layer $\epsilon_{1}$ will be 
close to a resonant TM $l=1$ permittivity value in order to generate this VSH (a resonant permittivity enables the existence of a field without a source as in a gain medium in laser). Similarly, all the other modes can be excited for the permittivity values close the eigenpermittivities, generating the time reversal of all the multipole radiation patterns, which correspond to all the emission and absorption transitions of atoms and molecules \cite{condon1934absolute}. Alternatively, a 
frequency which is close to an eigenfrequency can be used. Using an eigenpermittivity is advantageous in this context since the resonance can be fully reached by introducing a gain. While these eigenvalues
are usually associated with a gain that is needed to generate the field, there are some cases when they are real valued \cite{farhi2016electromagnetic} or have epsilon near zero \cite{alu2007epsilon,farhi2016electromagnetic}.

The electromagnetic field expansion for a physical electric field
$\boldsymbol{E}$ at a given angular frequency $\omega$ can be written as follows \cite{bergman1980theory} 
\begin{equation}
{\bf E}={\bf E}_{0}+\sum_{n}\frac{s_{n}}{s-s_{n}}\frac{\langle\tilde{{\bf E}}_{n}|{\bf E}_{0}\rangle}{\langle\tilde{{\bf E}}_{n}|{\bf E}_{n}\rangle}\left|{\bf E}_{n}\right\rangle ,
\end{equation}
where  $s_{n}\equiv\epsilon_{2}/\left(\epsilon_{2}-\epsilon_{1n}\right)$
is the eigenvalue, $\epsilon_{2}$ is the host-medium permittivity, $s\equiv\epsilon_{2}/\left(\epsilon_{2}-\epsilon_{1}\right),$
$\boldsymbol{E}_{n}$ and $\tilde{\boldsymbol{E}}_{n}$ are the eigenfunction
and its dual, and $\boldsymbol{E}_{0}$ is the incoming field. $\left\langle \boldsymbol{E}_{1}|\boldsymbol{E}_{2}\right\rangle =\int d\boldsymbol{r}\theta_{1}\left(\boldsymbol{r}\right)\boldsymbol{E}_{1}\cdot\boldsymbol{E}_{2}$
and $\theta_{1}\left(\boldsymbol{r}\right)$ is a window function
which equals 1 inside the inclusion volume. Thus, when $\epsilon_{1}$
is close to $\epsilon_{1n},$ $1/(s-s_n)\gg 1$ and the corresponding eigenfunction has a large contribution
in the electric field expansion (see for example Ref. \cite{farhi2017eigenstate}, Fig. 2). Clearly, other modes and the incoming
field exist in the expansion. Fortunately, close to a resonance, the
TM $l=1$ eigenfunction will have the dominant contribution inside the spherical volume.

Still, VSHs have a degeneracy in the $m$ index, which usually results
in the generation of all the $m$ modes as a response to an incoming
electric field. We therefore employ the current formulation of the
field expansion in order to remove this degeneracy. In this formulation
we express the incoming field in terms of Green's tensor ${\bf E}_{0}({\bf r})=\int dV\overleftrightarrow{G}\left(\mathbf{r},\mathbf{r}'\right)\cdot\mathbf{J}\left(\mathbf{r'}\right)$
and substitute it in $\langle\tilde{{\bf E}}_{n}|{\bf E}_{0}\rangle.$
Then, we change the order of integration and use the definition of
the eigenfunction to obtain \cite{farhi2016electromagnetic}
\begin{align}
\langle{\bf \tilde{E}}_{n}|{\bf E}_{0}\rangle=-\frac{4\pi i}{\epsilon_{2}\omega}\int dV'\theta_{1}\left(\mathbf{r}'\right){\bf E}_{n}\left(\mathbf{r}'\right)\cdot\int dV\overleftrightarrow{G}\left(\mathbf{r}',\mathbf{r}\right)\cdot\mathbf{J}\nonumber\\
=-\frac{4\pi i}{\epsilon_{2}\omega}s_{n}\int dV{\bf E}_{n}\left(\mathbf{r}\right)\cdot\mathbf{J}_{{\rm dip}}\left(\mathbf{r}\right)=-\frac{4\pi s_{n}}{\epsilon_{2}}\mathbf{p}\cdot\mathbf{E}_{n}({\bf r}_{0}),\nonumber
\end{align}
where $\mathbf{J}_{{\rm dip}}\left(\mathbf{r}\right)$ is an oscillating
point electric dipole, $\mathbf{p}$ is the dipole moment, and $\omega$
is the oscillation frequency. 

Now the expansion of the electric field reads 
\begin{align}{\bf E}={\bf E}_{0}-\frac{4\pi}{\epsilon_{2}}\sum_{n}\frac{s_{n}^{2}}{s-s_{n}}\frac{\mathbf{p}\cdot\tilde{{\bf E}}_{n}\left(\mathbf{r}_{0}\right)}{\langle\tilde{{\bf E}}_{n}|{\bf E}_{n}\rangle}\left|{\bf E}_{n}\right\rangle .\end{align}
Thus, situating an oscillating dipole may result in the generation of one TM $l=1$ mode
(see Fig. 1).

\begin{figure}
\includegraphics[scale=0.5]{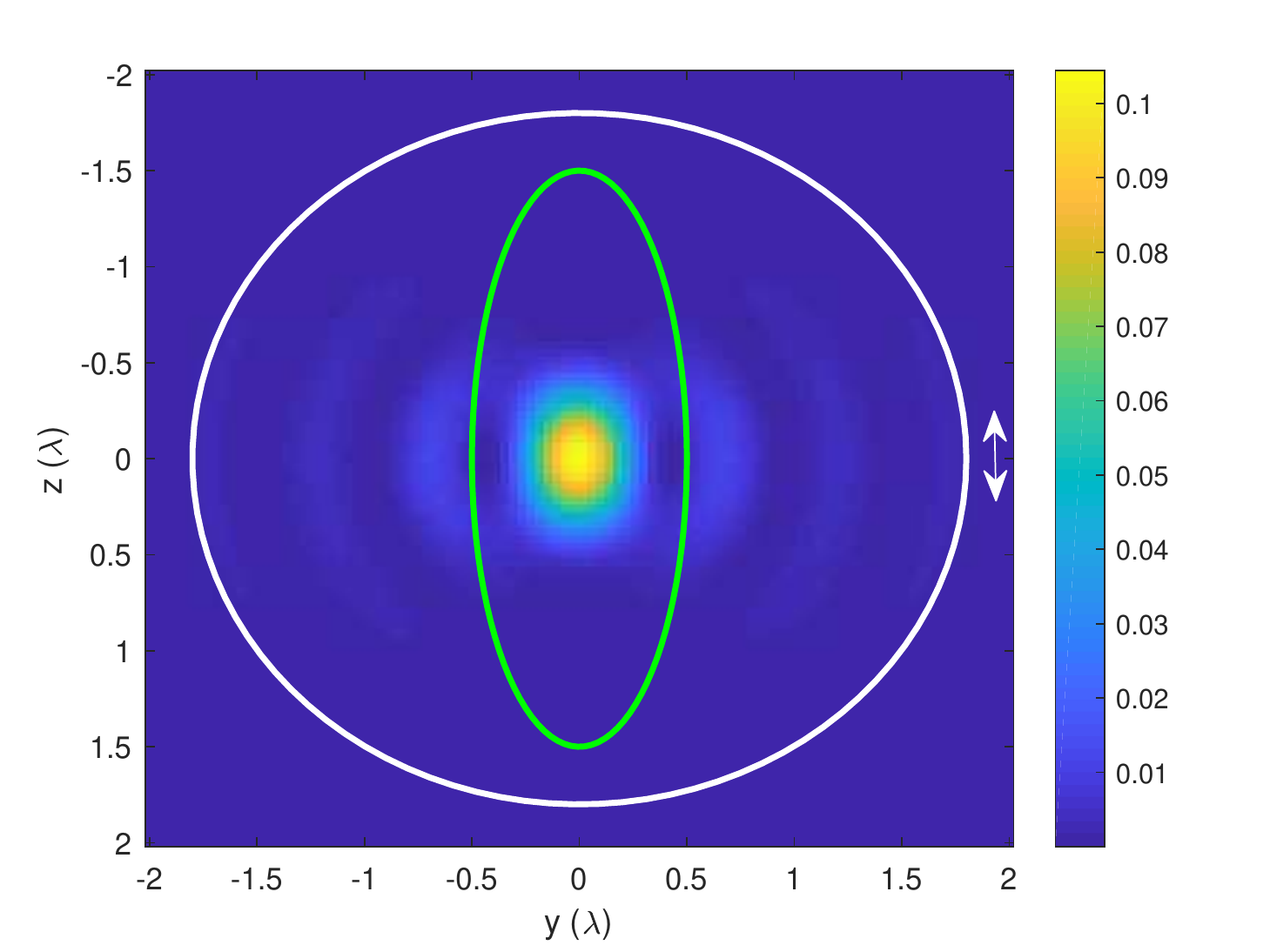}

\caption{A setup composed of a spherical layer (white) and an
oscillating dipole, and $|\mathbf{E}|^2$ of a TM $l=1$ eigenmode near the origin calculated analytically. The external ellipse
is the focal spot of an Airy disc (green). The size and thickness of the spherical
layer and the location of the oscillating dipole can vary. The spherical layer has a permittivity that is close to an eigenpermittivity $\epsilon_1\approx \epsilon_{1,\mathrm{TM}\, l=1}$ and the host-medium permittivity is $\epsilon_2=1.$ The intensity of the TM $l=1$ mode relative to the intensity of the incident field and the other modes depends on the proximity to the resonance.}
\end{figure}
The general form of a TM VSH is \cite{jackson2012classical}
\begin{align}
\boldsymbol{E}_{lm}^{^{\left(E\right)}}\propto&\frac{1}{\epsilon\left(r\right)}\nabla\times f_{l}\left(kr\right)\boldsymbol{X}_{lm},\,\,\,\boldsymbol{X}_{lm}=\frac{1}{\sqrt{l\left(l+1\right)}}\boldsymbol{L}Y_{lm},\nonumber\\
&f_{l}\left(kr\right)=A_{l}^{^{\left(1\right)}}h_{l}^{^{\left(1\right)}}\left(kr\right)+A_{l}^{^{\left(2\right)}}h_{l}^{^{\left(2\right)}}\left(kr\right),\nonumber
\end{align}
where $f_{l}\left(r\right)$ is a linear combination of spherical
Hankel functions, $h_{l}\left(r\right)$ is a spherical Hankel function,
$k$ is the wavevector, $\boldsymbol{L}=\frac{1}{i}\left(\boldsymbol{r}\times\nabla\right),$ and $Y_{lm}$ is a spherical harmonic.

For a spherical layer in $r_{1}<r<r_{2},$ $f_{l}\left(r\right)$
that satisfies boundary conditions is of the form
\[
f_{l}\left(r\right)=\left\{ \begin{array}{cc}
C_{l}h_{l}^{^{\left(1\right)}}\left(k_2 r\right) & r>r_{2}\\
B^{(1)}_{l}h_{l}^{^{\left(1\right)}}\left(k_{1n}r\right)+B_{l}^{^{\left(2\right)}}h_{l}^{^{\left(2\right)}}\left(k_{1n}r\right) & r_{1}<r<r_{2}\\
A_{l}j_{l}\left(k_2r\right) & r<r_{1}
\end{array}\right.,
\]
where $j_{l}\left(r\right)$ is a spherical Bessel function and $k_{1n},k_2$ correspond to $\epsilon_{1n},\epsilon_2,$ respectively. These eigenfunctions are standing waves for $r<r_1$ and propagating waves for $r>r_2$ at a given frequency. The eigenpermittivity $\epsilon_{1n}$ in  $r_{1}<r<r_{2}$ is calculated using an eigenvalue equation as we now explain.

An eigenpermittivity enables the existence of the field without a source and we therefore only need to impose boundary conditions.
From continuity of tangential $\boldsymbol{E}$ and $\boldsymbol{H}$
we have for a TM eigenfunction (assuming $\epsilon_2=1$)
\begin{align}
f_{\text{l}}\left(r_{1}^{-}\right)=f_{\text{l}}\left(r_{1}^{+}\right),\,\,\,f_{\text{l}}\left(r_{2}^{-}\right)=f_{\text{l}}\left(r_{2}^{+}\right),\nonumber\\
\left.\frac{\partial\left(rf_{\text{l}}\left(r\right)\right)}{\partial r}\right|_{r=r_{1}^{-}}=\frac{1}{\epsilon_{1n}}\left.\frac{\partial\left(rf_{\text{l}}\left(r\right)\right)}{\partial r}\right|_{r=r_{1}^{+}},\nonumber\\
\frac{1}{\epsilon_{1n}}\left.\frac{\partial\left(rf_{\text{l}}\left(r\right)\right)}{\partial r}\right|_{r=r_{2}^{-}}=\left.\frac{\partial\left(rf_{\text{l}}\left(r\right)\right)}{\partial r}\right|_{r=r_{2}^{+}},\nonumber
\end{align}
from which we obtain an eigenvalue equation and $\epsilon_{1n}.$
Similarly for a TE eigenfunction we write
\[
\boldsymbol{E}_{lm}^{^{\left(M\right)}}\propto f_{l}\left(kr\right)\boldsymbol{X}_{lm},
\]
with the boundary conditions
\begin{align}
f_{\text{l}}\left(r_{1}^{-}\right)=f_{\text{l}}\left(r_{1}^{+}\right),\,\,\,f_{\text{l}}\left(r_{2}^{-}\right)=f_{\text{l}}\left(r_{2}^{+}\right),\nonumber\\
\left.\frac{\partial\left(rf_{\text{l}}\left(r\right)\right)}{\partial r}\right|_{r=r_{1}^{-}}=\left.\frac{\partial\left(rf_{\text{l}}\left(r\right)\right)}{\partial r}\right|_{r=r_{1}^{+}},\nonumber\\
\left.\frac{\partial\left(rf_{\text{l}}\left(r\right)\right)}{\partial r}\right|_{r=r_{2}^{-}}=\left.\frac{\partial\left(rf_{\text{l}}\left(r\right)\right)}{\partial r}\right|_{r=r_{2}^{+}}.\nonumber
\end{align}
Clearly, the eigenpermittivities of the TE and TM modes depend on the radius and the thickness of the spherical layer.
%
%
%

The eigenfunctions in the radiation zone (far field) can be expressed as \cite{jackson2012classical}
\[
\boldsymbol{E}_{lm}^\mathrm{TM}\rightarrow Z_{0}\boldsymbol{H}_{lm}^{\mathrm{TM}}\times\boldsymbol{n},
\]
where $\boldsymbol{n}=\boldsymbol{r}/r.$ Hence, since $\boldsymbol{H}_{lm}^{\mathrm{TM}}\propto \boldsymbol{E}_{lm}^{\mathrm{TE}} $
is parallel to the sphere surface \cite{jackson2012classical}, $\boldsymbol{E}_{lm}^{\mathrm{TM}}$
is also parallel to the sphere surface. Thus, due to the inner product
in Eq. (2), when an oscillating dipole is placed in the radiation
zone it may excite a mode if it is oriented parallel to the spherical-layer surface.

For concreteness, we situate an oscillating dipole outside the spherical
layer on the positive $x$ axis. The $y,z$ components of $H_{lm}^{\mathrm{TM}}$ 
can be found from \cite{jackson2012classical}
\[
 H_{lm}^{\mathrm{TM}} \propto  \boldsymbol{L}Y_{lm},\,\,\,L_{y}=\frac{1}{2i}\left(L_{+}-L_{-}\right),\,\,L_zY_{lm}=mY_{lm}.
\]
The $z$ components of the TM $l=1$ eigenfunctions in the radiation zone readily follow from the
two relations above
\[
E_{l=1,m=0\,z}^{\mathrm{TM}}\neq0,\,\,\,E_{l=1,m=\pm1\,z}^{\mathrm{TM}}=0.
\]
Thus, by placing an oscillating dipole on the $x$ axis directed along
the $z$ axis we have removed the $m$ degeneracy of the TM modes. It can be seen that objects \emph{at all locations} will generate a focal spot at the origin. In addition, the $\theta$ dependency of the field can be written as $\boldsymbol{E}_{\mathrm{TM}\,l=1}\left(\theta\right)\propto\sin\theta\left(-\hat{x}\cos\theta+\hat{z}\sin\theta\right),$ which equals the $\theta$ dependency of the far field of an oscillating dipole and shows that the mode is indeed its time reversal. Oscillating dipoles on the $xy$ plane directed along $z$ will generate fields in the $z$ axis at the focal spot. From symmetry, situating several current sources will result in a superposition of the TM $l=1,m=0$ mode according to their locations and orientations. In addition, other forms of illumination (which correspond to current distributions) such as a laser illumination may also be used to generate a subwavelength focal spot (the current source may be associated with the gain medium).  
Also, since \cite{jackson2012classical}
$$ \sum_{m=-l}^{l}\left|\boldsymbol{X}_{lm}\left(\theta,\phi\right)\right|^{2}=\frac{2l+1}{4\pi}$$
 combining two spherical structures (e.g., a sphere and a spherical layer), each corresponding to a TM $l=1$ resonance at a given frequency, and using oscillating dipoles such that all the $m$ modes are excited, will result in isotropic radiation.

In order to have a dominant contribution of the TM $l=1$ modes, the physical
permittivity has to be much closer to the corresponding eigenpermittivity
compared with its distances from the eigenpermittivities of the other
modes. The high-order modes have a minor contribution to the expansion and
we can focus on a certain $l$ range when comparing these distances
\cite{bohren2008absorption}. 
The resonant permittivity usually has an imaginary part that corresponds
to gain. While incorporating gain in the spherical layer will bring
the system to a resonance, if a real-valued permittivity will be close
enough to a resonance, a similar effect is expected. The spacing between
resonances and the imaginary part of the permittivity depend on the
thickness of the spherical layer. A thin spherical layer will result
in a large eigenpermittivity gain and widely-spaced resonances. A
thick spherical layer will result in a a small imaginary part of the
eigenpermittivities and more closely spaced resonances. 

When the system is close to a resonance and there is a polarizable/absorbing medium at the origin, the dominant contribution to the electric field can be from the emission at the origin. 
From Eq. (2) it can be seen that the source location near the origin translates into $\mathbf{\tilde{E}}_n(\mathbf{r}_0)$ and the source magnitude is proportional to $\mathbf{E}_n(\mathbf{r}_0)$. We thus get that when the field is generated by the medium at the origin there is an additional factor of $\sqrt{2}$ in the effective FWHM in each dimension. This is demonstrated in Supplementary Material, Fig. 2.

An additional degeneracy arises when $r_{1},r_{2}\apprge10\lambda$
since at the $r\gg\lambda$ limit $j_{l},h_{l}^{\left(1\right)}$
have the form
\[
j_{l}\left(r\right)\rightarrow\frac{1}{r}\sin\left(r-\frac{l\pi}{2}\right),\,\,\,h_{l}^{\left(1\right)}\left(r\right)\rightarrow\left(-i\right)^{l+1}\frac{e^{ir}}{r}.
\]
As a result the even and odd eigenvalues will be almost identical.
A possible way to remove this degeneracy is to slightly change the
structure so that the eigenfunctions and the eigenvalues will change.
For example, the spherical layer can be capped from above (or in several places), which will
also enable to easily place objects inside. Alternatively, this high degeneracy can be utilized for a strong optical response of the system (e.g., strong scattering, enhancement of spontaneous emission etc.). 
This degeneracy is an asymptotic degeneracy and is in addition to the $m$ degeneracy so it includes a very large number of modes. In practice, an excitation at a given frequency can excite all the even/odd TE/TM modes. Similarly, such a degeneracy is also expected for a sphere inclusion and possibly cylindrical structures. Combining spherical structures may result in an all-mode degeneracy and further enhance the response of the system. Note that the total radiated power is a sum of the contributions of all the multipoles \cite{jackson2012classical}.

To cross validate our analysis we calculated for setups of spherical layers in vacuum the eigenmodes and $|\mathbf{E}|^2$ as a response to an excitation of a dipole and a current loop using Comsol. In Fig. 2 we present a TM $l=1$ mode for $\ensuremath{r_{1}=0.7,\,\,r_{2}=1.4\mu\mathrm{m}},\,\epsilon_1=1.5,\,\epsilon_2=1,\,\mathrm{and}\,\,\ensuremath{\omega_{\mathrm{TM}\,l=1}/2\pi=7.92781\cdot 10^{14}+7.397\cdot 10^7i},$ where $\omega_{\mathrm{TM}\,l=1}$ is an eigenfrequency. It can be seen that the focal-spot size (normalized by $\lambda$) matches the one in the analytical calculation presented in Fig. 1. Eigenmodes exist without a source, which in the eigenpermittivity formulation arises from gain in the spherical layer, similarly to a laser. In addition, $\omega_{l=1}$ is almost real and we therefore expect that at $\omega=\mathrm{Re(\omega_{l=1})},$ $\epsilon_{1l}\approx 1.5$ will be almost real. The eigenfrequencies in this case are closely spaced, which requires high precision in $\epsilon_1$ to obtain a resonance. We then considered a setup of $r_1=0.7,\,\,\,r_2=0.9  \mu \mathrm{m},\,\epsilon_2=1,\,$  $\lambda=430 \mathrm{nm}$ and an oscillating point dipole parallel to the spherical layer. We calculated $\epsilon_{\mathrm{TM}\,l=1}$ using the TM eigenvalue equation around $\epsilon_{1l}=1.5$ and substituted the result rounded to two digits after the decimal point as the physical permittivity $\epsilon_1$ in a Comsol simulation. In Fig. 3 we present $|\mathbf{E}|^2$ and $\mathbf{E}$ (arrows) for $\epsilon_1=1.75-0.7i,\,\,\mathbf{r}_0=1\hat{x}  \mu \mathrm{m}, \mathbf{p}=1\hat{z} \mathrm{mA}$ in axial cross section. It can be seen that the focal-spot normalized size matches the ones in Figs. 1 and 2. Situating the dipole at any other distance will also result a focal spot at the origin, unlike imaging using a lens. In Fig. 4 we present $|\mathbf{E}|^2$ for a setup with a current loop with $\epsilon_1=1.45 -0.57i, \epsilon_2=1,  r_1=1.7, r_2=2\mu \mathrm{m},\lambda=430 \mathrm{nm}, \mathbf{J}=1\hat{z}\mathrm{A/m^2},\, \mathrm{and}\, r_0=2.2\mu \mathrm{m}.$ Interestingly, the field intensity is much stronger at the origin compared to the one around the current loop. In addition, the current distribution reminds a gain medium distribution in a laser, which may mean that a laser can also be used to generate this TM $l=1$ mode. In the Supplementary Material we demonstrate focusing using a capped spherical layer with $r_2\approx 10\lambda$. This structure has a full azimuthal-angle coverage unlike focusing light using two lenses. In all the simulations we used a perfectly matched layer to account for boundary conditions (external layer).
 
Now we analyze the TM eigenpermittivities for a sphere inclusion in vacuum as we increase the sphere radius. Similarly to the spherical-layer setup all the odd/even eigenvalue equations coalesce when increasing the sphere radius $r_1.$ In Fig. 5 we present $\epsilon_{1l\,\mathrm{TM}}$ as a function of $r_1$ for $\lambda=430\mathrm{nm.}$ The eigenpermittivities have a negligible imaginary part (smaller than $10^{-8}$) and we therefore present only the real part. It can be seen that for $r_1>8\mu \mathrm{m}$ all the even/odd eigenvalues are practically the same. Thus, using a physical permittivity $\epsilon_1$ that is close to the odd or even eigenpermittivity, will excite all (or most) of these eigenstates, leading to a very strong response of the system (without requiring gain in this case). Note that at large sphere radii the eigenvalues are more robust to changes in the radius.

We now evaluate the enhancement of various optical phenomena due to the infinite-asymptotic degeneracy.
We investigate the enhancement of spontaneous emission \cite{spontaneous2015} of a dipole in a sphere/spherical-layer setup when $r_1,r_2,r_{\mathrm{dipole}}\gg\lambda$ due to the infinite degeneracy. To that end, we write the expression for the density of states  \cite{Taflove2013,Wijnands1997,Lagendijk1996}, which is dominant in Fermi-Golden-Rule calculation \cite{cohen1977quantum}
$$\rho_\mu=-\frac {2\omega}{\pi}\mathrm{Im}\left[G_{\mu\mu}(\mathbf{r},\mathbf{r}',\omega)\right].$$ 
We then evaluate the sum in the eigenfunction expansion in Eq. (2). We consider a spherical layer with a physical permittivity that is slightly above the first or second eigenpermittivity, namely $\epsilon_1>\epsilon_{11}$ or  $\epsilon_1>\epsilon_{12}$ (see Fig. 5). In this situation $s_l^2/(s-s_l)$ have the same sign and approximately \emph{the same value} for a very large number of modes (e.g., at least 20 modes for $r_1,r_2\geq40\lambda).$ We now analyze ${\tilde{E}}_{l\mu}(\mathbf{r}_{\mathrm{dipole}}){{E}}_{l\mu}(\mathbf{r}_{\mathrm{dipole}}).$ Since the dual eigenfunctions \cite{bergman1980theory} 
$$\boldsymbol{\tilde{E}}_{lm}^{\left(E\right)}\propto\nabla\times f_{l}\left(kr\right)\boldsymbol{X}^*_{lm},\,\,\, \boldsymbol{\tilde{E}}_{lm}^{\left(M\right)}\propto f_{l}\left(kr\right)\boldsymbol{X}^*_{lm},$$ we get that the phase of ${\tilde{E}}_{l\mu}(\mathbf{r}_{\mathrm{dipole}}){{E}}_{l\mu}(\mathbf{r}_{\mathrm{dipole}})$ is determined by $f_l.$ Since $ f_{l}\approx  f_{l+2}$ we get approximately the same phase for all the modes whose eigenvalues coalesce. Similar arguments apply for the inner product $\langle\tilde{{\bf E}}_{n}|{\bf E}_{n}\rangle$, see Appendix A in Ref. \cite{bergman1980theory}. For example, the integral in the inner product of the TE modes can be performed analytically and can be shown to be invariant to $l\rightarrow l+2.$ This leads to a constructive interference in the field summation. Thus, if we have $n$ modes that have effectively the same eigenvalue, their resonance constributions will add constructively and we get approximately $n$ times enhancement in the density of states compared with a standard resonance of the same structure. Clearly, the larger the spherical-structure radius, the more eigenvalues will be effectively the same (see Fig. 5). This should be multiplied by the enhancement factor that arises from the proximity of the physical permittivity to the resonant permittivity $(\propto1/(s-s_n)$ from Eq. (2)). See Ref. \cite{farhi2017eigenstate} in which the modes also interfere constructively. In this reference when the physical permittivity is close to first eigenpermittivity $s_l^2/(s-s_l)$ decays upon increasing $l$ and when the physical permittivity is close to the accumulation point the field is enhanced very close to the metal-dielectric interface since the high-order modes, which have approximately the same $s_l^2/(s-s_l),$ decay spatially rapidly. Here the modes have approximately the same $s_l^2/(s-s_l)$ contribution and they all scale as $1/r$ at large distances, leading to a strong response that extends relatively far from the dielectric sphere.

We proceed to analyze the enhancement of absorption and stimulated emission induced by a dipole on itself due to the infinite degeneracy. Clearly, the enhancement of the density of states in Fermi-Golden-rule calculation \cite{cohen1977quantum} will be the same. Assuming that the multiple expansion for light-matter interaction holds and the dipole interaction is the dominant interaction for all the modes in the field expansion, we get that if $n$ modes are effectively on resonance and the field that is generated by the dipole is enhanced by a factor of $n,$ $\left|\left\langle \psi_{j}|H_{\mathrm{int}}|\psi_{i}\right\rangle \right|^{2}$  will scale as $n^2$ and the overall enhancement will scale as $n^3$ compared with a standard resonance. This is a very large enhancement and for $n=20$ we get an enhancement factor of 8000  (needs to be multiplied by $\propto1/(s-s_n)^3$). Note that a sphere with $r_1=20\lambda$ is of the order of a human cell for visible and infrared light and hence this phenomenon has potential use in biomedical applications such as targeting cells with light. Another potential application is omnidirectional antenna/detector, which directs its its field pattern according to the source location.

The enhancement of the scattering arises from the fact that the total radiated power is a sum of the contributions of all the multipoles \cite{jackson2012classical}. Thus, we deduce that the total power is enhanced by a factor of $n,$ compared with a standard resonance, where $n$ is the number of modes that are effectively on resonance. This should be multiplied by the enhancement factor that arises from the proximity of the physical permittivity to the resonant permittivity $\propto 1/(s-s_n)^2.$ Similar analysis follows for the enhancement of  absorption by a sphere as the absorption power is given by $\omega\cdot\mathrm{Im}(\epsilon_1)|\mathbf{E}|^2/2$ \cite{jackson2012classical} and many modes can be excited inside the sphere.

\begin{figure}
\includegraphics[scale=0.5]{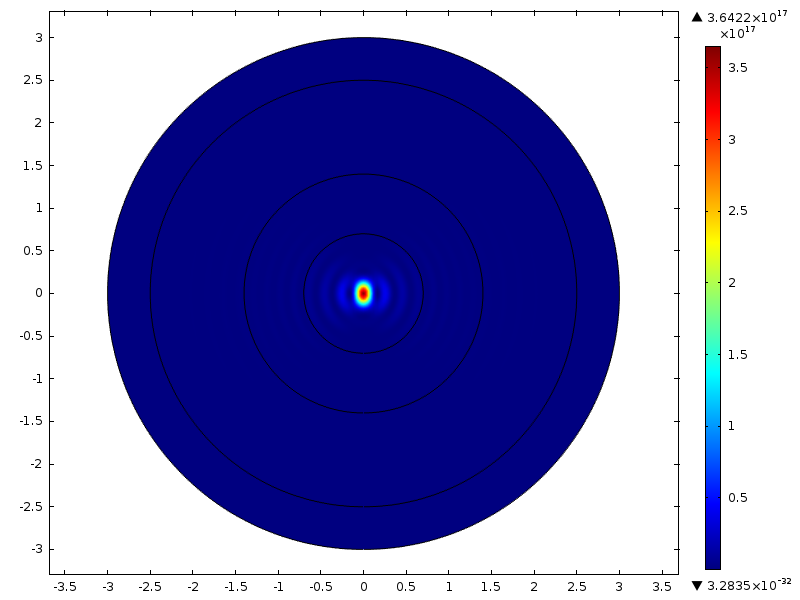}

\caption{TM $l=1$ eigenmode for a setup of a spherical layer in vacuum with $\ensuremath{r_{1}=0.7,\,\,r_{2}=1.4\mu\mathrm{m}},\,\epsilon_1=1.5,\,\mathrm{and}\,\,\ensuremath{\omega_{l=1}/2\pi=7.92781\cdot 10^{14}+7.397\cdot 10^7i}.$ }
\end{figure}

%

\begin{figure}

           \centering
        \includegraphics[width=8cm]{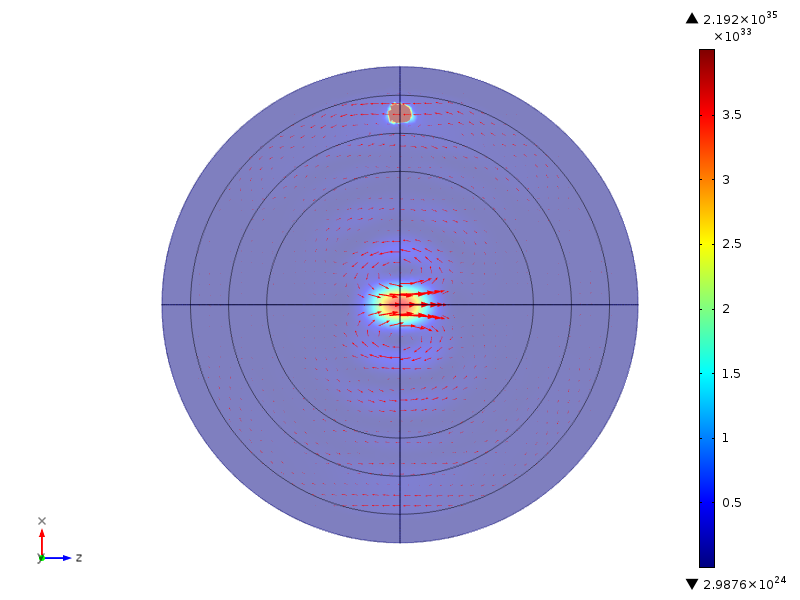}
      
    \caption{$|\mathbf{E}|^2$ and $\mathbf{E}$ (arrows) for a setup of a spherical layer in vacuum and an oscillating point dipole with $r_1=0.7,\,\,\,r_2=0.9  \mu \mathrm{m},\epsilon_1=1.75-0.7i,\,\,\mathbf{r}_0=1\hat{x}  \mu \mathrm{m}, \mathbf{p}=1\hat{z} \mathrm{mA}$ and $\lambda=430 \mathrm{nm}.$ The field intensity near the dipole is capped and the value on top of the bar is the maximum value in the calculation.}
\end{figure}

\begin{figure}
\includegraphics[scale=0.5]{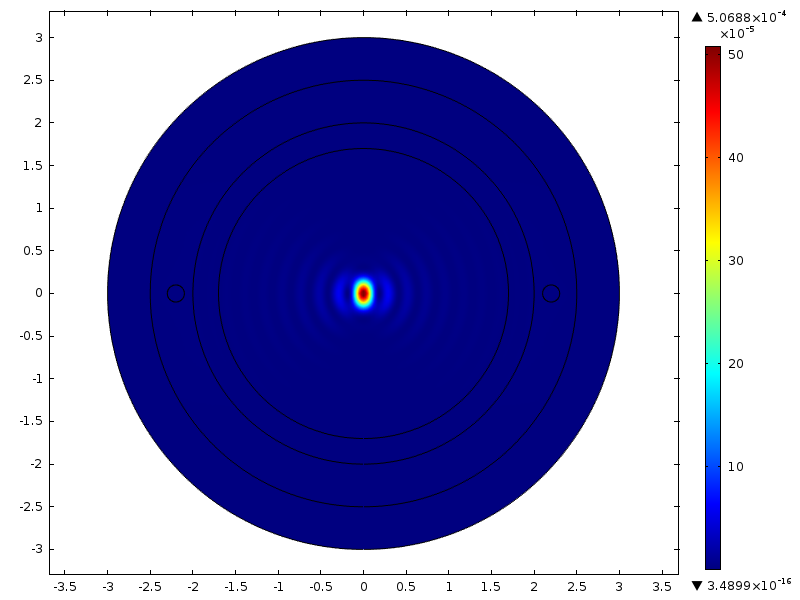}

\caption{$|\mathbf{E}|^2$ for a setup of a spherical layer in vacuum with a current loop and $\epsilon_1=1.45 -0.57i, r_1=1.7, r_2=2\mu \mathrm{m},\lambda=430 \mathrm{nm}, r_0=2.2\mu \mathrm{m}.$}
\end{figure}

\begin{figure}
\includegraphics[width=9cm]{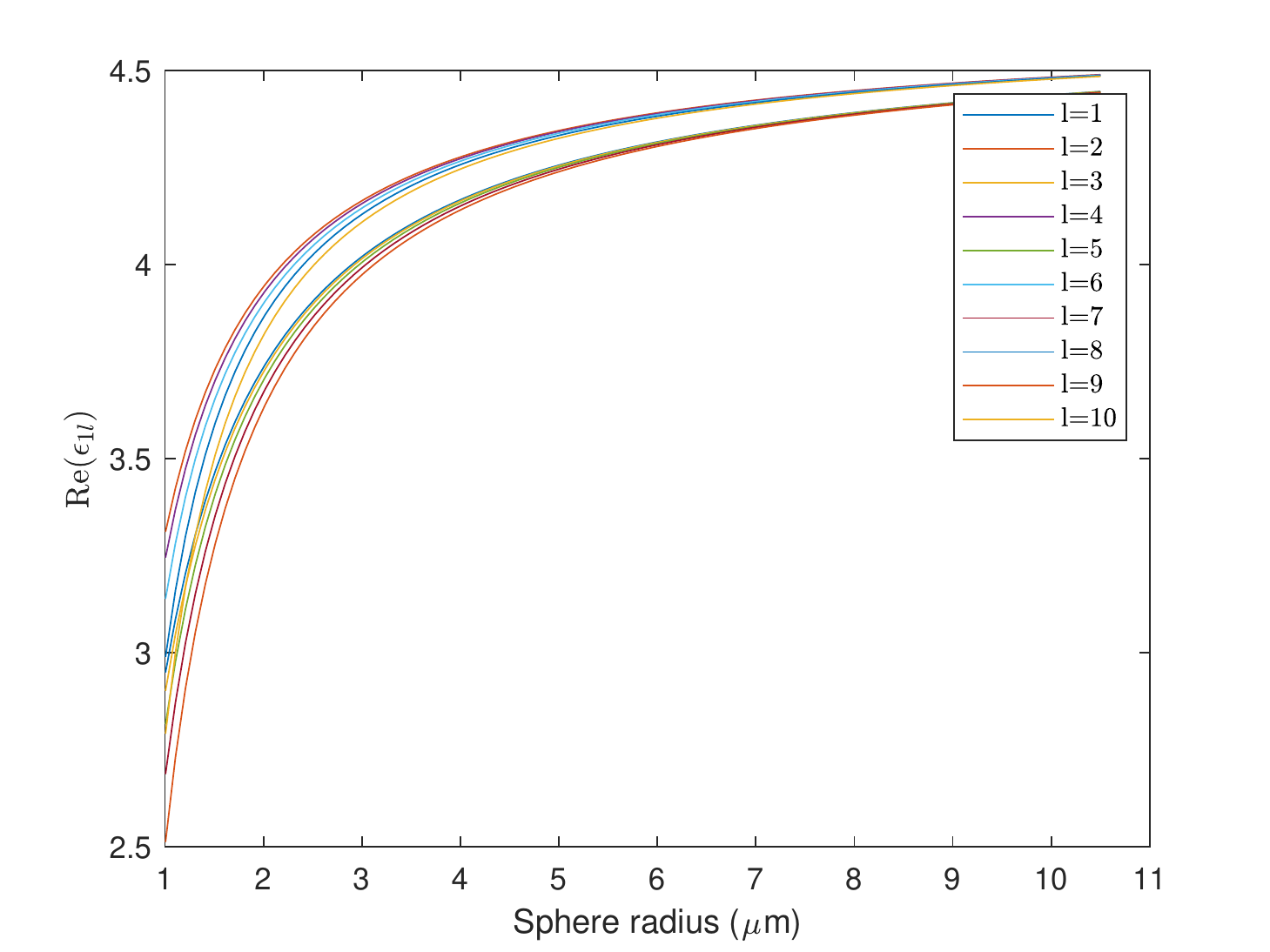}

\caption{ $\epsilon_{1l\,\mathrm{TM}}$ as a function of the sphere radius for a sphere in vacuum at $\lambda=430$nm}
\end{figure}

We now explore two directions to localize electric field with deep subwavelength resolution.
We first present a three-body-resonance mechanism in which we slightly change the permittivity value of the spherical layer to move the system away from resonance and introduce a spherical particle that will bring the system back to resonance when located at the origin.
We consider a spherical layer in a host medium that is off-resonance and close to a resonance, possibly having a dielectric material with gain.
We then introduce a spherical particle that when situated at the origin results in a TM $l=1$ resonance of the three-body system for a given permittivity value of the particle, possibly a dielectric material that is different from the host-medium permittivity, and we set the physical permittivity value of the particle to be equal to this eigenpermittivity.
For a different location of the particle the system will be on resonance for a different permittivity value of the particle.
This setup translates location changes of the particle to changes in the eigenpermittivity, utilizing the $1/(s-s_n)$ factor to localization of the particle.
We thus may achieve strong localization capability of the system - for a slight change in the location of the particle the field intensity everywhere will change significantly. To translate this idea into practical applications one can use frequencies for which the host medium that can have in general spatially-varying permittivity, is relatively uniform/transparent.

We also consider the possibility that time reversal of the field emitted in a multipole transition at the origin of an atom or molecule will spatially match the quantum transition current. It was suggested based on a classical wave equation analysis that when the time reversal of the field emitted by a point source impinges on a perfect absorber at the origin, the field pattern will have a $1/r$ scaling near the origin \cite{ma2018towards}. Now we turn to the quantum analysis. We first note that in the semi-classical quantum treatment in Ref. \cite{condon1934absolute} there is a $1/r$ scaling in the transition-rate calculation. One can think that the time reversal of an emission process is absorption, having a field with a $1/r$ dependency near the origin. In practice, emission and absorption are related to the transition between electronic or nuclear eigenstates. We can thus expect that the field will not diverge and think of a classical analogue of a dipole with a characteristic size of the average distance of the probability density function from the center of mass. Let us analyze the emission process and its time reversal. We consider a hydrogen atom for simplicity and assume that there is a transition from an eigenstate $\psi_1$ to an eigenstate $\psi_2.$ As a result of the spatial change in the probability density function electric field is emitted. We express the quantum current $\mathbf{j}=\mathbf{v}\rho=\mathbf{v}|\psi|^2=\frac{1}{2m}\left(\psi^{*}\boldsymbol{p}\psi-\psi\boldsymbol{p}\psi^{*}\right),$ where $\psi$ is the wavefunction that can transition between states and $\mathbf{v}$ is the group velocity of the particle \cite{cohen1977quantum}. The electric field $\mathbf{E}$ then propagates in space occupying a spherical shell. Now we time reverse the process. We assume that the field is generated on the spherical shell. The field then propagates toward the atom/molecule. We assume that when the field reaches the atom or molecule they are in the same state as when they emitted the field up to a $\pi$ phase difference in $\mathbf{v}$. Using reciprocity and treating the quantum current as classical the field near the origin will then be in the same form of the quantum current $\mathbf{j}$ that generated the field. Thus, the field pattern matches the form of the transition current and can be optimal for driving the transition. The field is thus deep subwavelength with typical size of the average distance of the density function from the center of mass. In this situation the spatial variations of the electric field are comparable to the electron/nuclear wavefunction and the spatial variations of $\mathbf{E}$ or $\mathbf{A}$ will have to be taken into account explicitly in light-atom/molecule interaction calculations. In standard light-atom/molecule interaction, the term $-\frac{q}{m}\mathbf{P}\cdot\mathbf{A}$ for the value of $\mathbf{A}$ at the atom/molecule location drives the dipole transition. However, $\mathbf{A}$ is constant \cite{cohen1977quantum} and not necessarily spatially overlaps optimally with the current that drives the transition. This absorption process can be complemented by stimulated emission for a field that oscillates at a frequency $\omega.$ Note that the resonant spherical layer should be tuned to this $\omega.$ This process can have unique characteristics such as strong absorption and emission, high-order multipole transitions involved etc.
See for example Refs. \cite{Agostini79,Corkum2007,Ghimire2019,Dahan2019} in which phase matching of the electric field to the electron wave function results in a stronger interaction.
 If this is indeed the situation it would make sense that a slight change in the position of the atom/molecule from the origin will bring the interaction to the standard multipole-expansion interaction. Hence, if this can be realized experimentally using a resonant spherical layer, it may enable to localize atoms/molecules with deep-subwavelength resolution (for scattering medium with ballistic photons). In addition, we note that this description is applicable to all the transition types (dipole, quadrupole etc.).
While it is true that the spontaneous-emission rate of high-order radiation multipoles is usually slow, when it will occur for an atom/molecule at the origin the incoming time-reversed field can spatially match the transition current and drive the transition. 
Alternatively, transitions can be driven by an external current source. In order for the spherical layer to respond to several transitions one can utilize the infinite-asymptotic degeneracy.  
Note that when radiation is emitted by an atom/molecule at the origin, the spherical-layer setup generates the time-reversed field also according to the orientation, which maximizes the spatial overlap when interacting with the atom/molecule.

In addition, close to a resonance the density of states given by $\rho_\mu=-\frac {2\omega}{\pi}\mathrm{Im}\left[G_{\mu\mu}(\mathbf{r},\mathbf{r}',\omega)\right]$ \cite{Taflove2013,Wijnands1997,Lagendijk1996}, where $G_{\mu\mu}(\mathbf{r},\mathbf{r}',\omega)$ can be expressed as the electric field due to a dipole at the dipole location and direction in Eq. (2). Thus, when approaching a resonance the density of states and the field increase and as a result the transitions are enhanced. Hence, quantum mechanically we have enhancement in two aspects: field overlap with the transition current and increase in the density of states and electric field.  

We introduced a setup of a spherical layer, that close to a resonance
 generates the time reversal of the atomic and molecular multipole transitions. The time reversed signal in our setup is spatially continuous and is naturally generated by a medium with a uniform permittivity. 

We started by situating the spherical layer in a uniform medium, which generates a subwavelength free-space focal spot in three dimensions. The degeneracy of the excited mode is removed by incorporating currents on a plane
which is perpendicular to the spherical layer. Such currents can be realized by a medium, which is polarized due to an impinging electric field or even a laser source. Interestingly, when situating an object at the origin the field emitted by the polarized medium at the focal spot excites the TM $l=1$ spherical layer mode, which reexcites the medium at the focal spot etc. This coupling can enhance the emission from the medium at the focal spot. Also, near a resonance the field becomes very strong and may enable larger penetration of ballistic photons and enhancement of the signal generated at the focal spot by the spherical layer. To image from the focal spot, one can think of collecting light from the other side of the spherical layer by means of a lens or another optical element. This signal is mostly composed of the sum of the excitation of the TM $l=1$ mode due to the sources and the polarized medium at the focal spot, which may enable to acquire also the phase in the measurement. To further minimize the effective focal-spot size techniques such as nonlinear optics, PALM or STORM  \cite{betzig2006imaging,moerner1989optical}, and quantum imaging \cite{tenne2018super} can be used. In addition, the TM $l=2$ and TE $l=1$ modes have a torus shape \cite{farhi2017generating} and may be used to stimulate fluorescence emission at another wavelength similarly to STED \cite{hell1994breaking}.

We then explored two directions to localize field with deep subwavelength resolution. We presented a three-body-resonance mechanism in which we slightly change the permittivity value of the spherical layer to move the system away from resonance and introduce a spherical particle that will bring the system back to resonance when located at the origin. We then situated an atom or molecule at the origin and considered the possibility that the time reversed field of a multipole transition will generate field near the origin that spatially correlates with the quantum-transition current, resulting in a much stronger interaction at the origin.

The resonant spherical shell setup differs from a spherical cavity in several aspects: 1. It enables light from outside of the spherical shell to generate field inside and vice versa. 2. There is a strong amplification of the signal. Thus, even spontaneous emission can generate substantial field at the focal spot. When the system is on resonance, the mode is generated without a source. 3. It couples to a single multipole or equivalently an atomic/molecular transition spatially and temporally. 

This analysis is applicable to all wavelengths and due to its wave nature it may also apply to acoustics, in which gain materials were recently introduced  \cite{willatzen2014acoustic}, and matter waves. In addition, each spherical-layer mode has several eigenvalues and therefore there is flexibility in choosing the spherical-layer material, which may have importance for frequencies where it is more challenging to find materials that can focus waves \cite{graydon2018tight}. Importantly, it was shown that spherical waves (VSHs) can be generated by a single source, which may enable their practical generation, also at high frequencies where current modulation is impractical. Potential applications  are high-resolution 3D imaging and precise tissue ablation. In addition, the fact that this setup has a very high Q factor may be utilized to cavity QED, entanglement, and quantum information \cite{raimond2001manipulating}. Finally, for spherical structures with $r_{1}\apprge10\lambda$ there are all-odd and all-even TM/TE eigenvalue degeneracies, which results in a variety of optical phenomena of the system close to one of these eigenvalues. Combining spherical structures e.g., a sphere and spherical layer(s), each with a permittivity close to one of these resonances, may even result in an all-mode resonance.

\subsection*{Acknowledgments}
M. Segev is acknowledged for the funding. G. Bartal, D. Oron, and I. Kaminer are acknowledged for the useful comments.

\bibliographystyle{plain}

\end{document}